\documentclass[sigconf]{acmart}
\AtBeginDocument{%
  }

\setcopyright{acmlicensed}
\copyrightyear{2018}
\acmYear{2018}
\acmDOI{XXXXXXX.XXXXXXX}
\acmConference[Conference acronym 'XX]{Make sure to enter the correct
  conference title from your rights confirmation email}{June 03--05,
  2018}{Woodstock, NY}
\acmISBN{978-1-4503-XXXX-X/2018/06}

\usepackage{graphicx}
\usepackage{multirow}
\usepackage{tabularx}
\usepackage{booktabs}
\usepackage{makecell}   
\usepackage{array}    

\begin{document}

\title[Mitigating Silent Driving System Failures through Prospective Situation Awareness Enhancing Interfaces]{From Awareness to Intent: Mitigating Silent Driving System Failures through Prospective Situation Awareness Enhancing Interfaces}

\author{Jiyao Wang}
\authornote{Both authors contribute equally.}
\email{jwanggo@connect.ust.hk}
\orcid{0000-0002-0743-0121}
\affiliation{%
  \institution{The Hong Kong University of Science and Technology (Guangzhou)}
  \city{Guangzhou}
  \state{Guangdong}
  \country{China}
}

\author{Song Yan}
\email{syan931@connect.hkust-gz.edu.cn}
\authornotemark[1]
\affiliation{%
  \institution{The Hong Kong University of Science and Technology (Guangzhou)}
  \city{Guangzhou}
  \state{Guangdong}
  \country{China}
}

\author{Xiao Yang}
\email{xyang856@connect.hkust-gz.edu.cn}
\orcid{0000-0001-6096-9903}
\affiliation{%
  \institution{The Hong Kong University of Science and Technology (Guangzhou)}
  \city{Guangzhou}
  \state{Guangdong}
  \country{China}
}

\author{Qihang He}
\email{e1538701@u.nus.edu}
\orcid{0009-0003-4134-5291}
\affiliation{%
  \institution{National University of Singapore}
  \country{Singapore}
}

\author{Ange Wang}
\email{awang324@connect.hkust-gz.edu.cn}
\orcid{0009-0004-6175-5631}
\affiliation{%
  \institution{The Hong Kong University of Science and Technology (Guangzhou)}
  \city{Guangzhou}
  \state{Guangdong}
  \country{China}
}

\author{Chenglin Liu}
\email{cliu549@connect.hkust-gz.edu.cn}
\orcid{0009-0002-3568-6439}
\affiliation{%
  \institution{The Hong Kong University of Science and Technology (Guangzhou)}
  \city{Guangzhou}
  \state{Guangdong}
  \country{China}
}

\author{Chenglin Chen}
\email{cchen363@connect.hkust-gz.edu.cn}
\orcid{0009-0008-6281-409X}
\affiliation{%
  \institution{The Hong Kong University of Science and Technology (Guangzhou)}
  \city{Guangzhou}
  \state{Guangdong}
  \country{China}
}

\author{Zhenyu Wang}
\email{zwang209@connect.hkust-gz.edu.cn}
\affiliation{%
  \institution{The Hong Kong University of Science and Technology (Guangzhou)}
  \city{Guangzhou}
  \state{Guangdong}
  \country{China}
}

\author{Dengbo He}
\email{dengbohe@hkust-gz.edu.cn}
\orcid{0000-0003-4359-4083}
\authornote{Corresponding author}
\affiliation{%
  \institution{The Hong Kong University of Science and Technology (Guangzhou)}
  \city{Guangzhou}
  \state{Guangdong}
  \country{China}
}

\renewcommand{\shortauthors}{Wang et al.}

\begin{abstract}
  Silent automation failures, where a system fails to detect a hazard without warning, pose a critical safety challenge for partially automated vehicles. While research has mostly focused on takeover requests, how to support a driver in silent failure remains underexplored. We conducted a multi-modal driving simulator study with 48 participants to investigate how different Prospective Situation Awareness Enhancement (PSAE) interfaces, delivered via augmented reality head-up display, affect takeover performance. By integrating behavioral, subjective psychological, and physiological data, our analysis suggests that situational awareness (SA) serves as an important moderating factor through which PSAE interfaces improve takeover performance. Further, we found that providing perceptual cues was most effective in enhancing SA, while communicating system intent was superior for building trust. Finally, we identified a potential correlate of SA in the neuroactivity. Overall, this paper contributes to understanding how transparency-oriented interfaces may support drivers and provides design insights into HMI design for silent failures.
\end{abstract}


\begin{teaserfigure}
  \includegraphics[width=\textwidth]{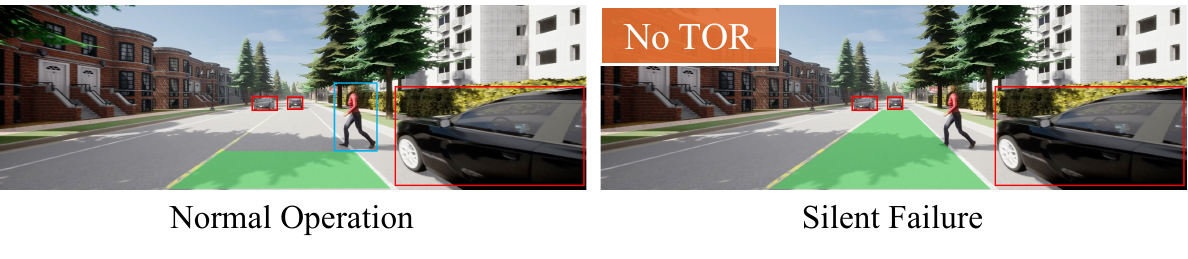}
  \caption{Illustration of the difference between an automated driving system (ADS) with normal operation and the silent failures. In case of failures, the ADS continues to control the vehicle, but its internal representation of the situation is misaligned with the actual road environment: either the hazard is not detected at the perception layer and/or the planned trajectory conflicts with the hazard at the planning/decision layer, and no TOR is issued in case of all failures.}
  \label{normal_silent}
\end{teaserfigure}


\maketitle

\section{Introduction}

The increasing integration of automation into daily life fundamentally changes the nature of human-technology interaction \cite{hancock2019future}. A core challenge in this evolving relationship is designing for transparency and trust in a shared-control paradigm, especially when a human operator must remain "in the loop." In the context of partially automated vehicles (SAE Level 2) \cite{on2021taxonomy}, which are gradually dominating the consumer market \cite{highway2023predicted}, this is not just a user experience problem but a critical safety issue. At this level, human drivers are still responsible for monitoring the environment and responding to system limitations, either actively or prompted by takeover requests (TORs) from automated driving systems (ADS). 

Existing human-ADS interaction research predominantly focused on TOR-initiated takeover scenarios \cite{wang2025exploring,srour2024s,shi2024effects,huang2025effect}. However, silent failures without TORs are likely more common and safety-critical in SAE Level 2 vehicles. Hazard denotes any on-path road user or object whose projected motion would conflict with the ego vehicle in the absence of driver intervention \cite{yan2024ch}. As shown in Figure \ref{normal_silent}, the inability of ADS to detect and respond to road hazards poses a severe challenge that has not yet been fully addressed in existing research. In these highly safety-critical situations, a system may encounter a failure without warning drivers in advance or until very late (e.g., due to an unforeseen malfunction), leaving the driver to detect and respond to the failure independently with a very short time buffer \cite{kanaan2024automation}. Additionally, the time budget and cognitive workload needed from drivers when they encounter hazards are shaped by how perceptually accessible those hazards are \cite{sun2019effects}. Previous studies \cite{yan2024ch,crundall2016hazard} distinguish two scenario types based on hazard visibility: visible hazards, which are fully observable as they materialize, and invisible hazards, which are initially hidden and must be detected via contextual cues. Invisible hazards typically demand earlier predictions and broader visual searches; while in ADS with silent failures, these differences in hazard visibility translate directly into distinct situation awareness requirements and takeover time budgets.

Some works \cite{goodge2024can,lee2024designing,kim2023and} have investigated the use of explanations to enhance driver awareness, but these works assume the system’s internal perception is aligned with the actual environment. In fact, this assumption breaks down in silent failure scenarios, where a malfunction in system perception can be the cause of system failure \cite{cummings2014man}. Despite a few studies \cite{jung2023projecting, feierle2022augmented} have explored improving takeover performance in silent failure with different explanatory information, to our knowledge, little research has investigated how information regarding key ADS components, i.e., perception and maneuver planning, influences drivers’ takeover decisions in silent failure events under different environmental conditions (e.g., lighting and hazard types), which would manipulate the complexity and difficulty of the event \cite{huang2024enhancing}.

To address this gap, we investigate the efficacy of the prospective situation awareness enhancements (PSAE) \cite{gregoriades2016enhancing} in silent failure scenarios. This form of information, fundamentally different from retrospective explanations that justify past system actions after a failure has occurred, is designed to enhance the driver’s Level 1 (\textit{Perception}) and Level 3 (\textit{Projection}) SA \cite{endsley2017toward} by offering a continuous, forward-looking stream of information about the system’s perceptual state (what the system ‘sees’) and planned maneuver. Drawing on theories of embodied cognition and cognitive neuroscience \cite{clark2013whatever}, we posit that a driver’s takeover behavior is not merely a reactive response but the culmination of a dynamic interplay between their psychological state, neuro-physiological activities, and the environmental context \cite{hancock2020imposing}. A driver’s ability to respond to a hazardous event is contingent on the timely and coordinated interplay of mental processing and motor preparation \cite{korber2016influence}. In line with classical work on driving as hierarchical information processing \cite{michon1985critical} and with situation awareness theory \cite{endsley1995measurement}, we assume that drivers must first perceive, comprehend, and project the state of the environment before they can select and execute appropriate actions \cite{markkula2018models,gonccalves2019applicability,thomas2021uncovering}. In addition, indicators from neuro-physiological signals can provide continuous, objective correlates of attentional engagement and cognitive activities \cite{mehta2013neuroergonomics}. 

Therefore, based on dynamic augmented–reality head–up display (AR-HUD), we adopt a multi-layered framework, in which PSAE as an experimental variable, to empirically examine how its variation, in conjunction with ambient environmental factors, including lighting conditions (day vs. night) and hazard visibility (visible vs. invisible), influences drivers’ SA, perceived safety, trust, EEG/ERSP and EMG-based measures, and ultimately their takeover performance \cite{de2014effects}. This provides a more nuanced understanding of the perception and trajectory planning information of ADS may affect human-ADS cooperation in safety-critical scenarios. Specifically, we use electroencephalography (EEG) to analyze Event-Related Spectral Perturbations (ERSP) \cite{makeig1993auditory}, a measure of oscillatory brain activity associated with cognitive processes, and electromyography (EMG) to track motor preparation \cite{coles1989modern}. From these, we measure the temporal alignment between the peak of a driver's cognitive processing and the onset of their motor preparation, thereby providing a unique insight into the temporal coupling of cognition and action \cite{pfurtscheller1999event} when drivers are provided different PSAE information in different silent failure scenarios.

To investigate these complex relationships and address the aforementioned research gaps, this study aims to answer the following research questions:

\begin{itemize}
\item \textbf{RQ1}: How do different PSAE designs and environmental factors (i.e., lighting and hazard visibility) influence drivers' self-reported Situation Awareness (SA), perceived safety, and trust, as well as objective EEG- and EMG-derived physiological indicators during silent automation failures?
\item \textbf{RQ2}: What is the intricate relationship between a driver's subjective psychological states (e.g., SA, perceived safety, trust) and their physiological indicators?
\item \textbf{RQ3}: What are the underlying mechanisms through which PSAE interfaces impact takeover performance? Specifically, do psychological states and/or physiological indicators serve as the primary causal pathways?
\end{itemize}

Our contributions are as follows: (1) We quantitatively model the multifaceted impacts of PSAE interfaces, lighting, and hazard visibility on drivers' internal states (i.e., psychological states and physiological indicators) in the context of silent ADS failures. (2) We examine hierarchical pathways connecting driver psychology and physiology to takeover behavior, and our models suggest that, in this study, the effect of the interface can be indirectly mediated by drivers’ SA. (3) We offer an initial mechanistic interpretation of how our intervention may influence safety-related outcomes in our simulator setting, and derive empirically grounded design implications for creating transparent and trustworthy automated systems in silent failures.

\section{Related Work}

\subsection{Silent Failures in Automated Driving}
SAE Level~2 places human drivers in a critical supervisory role, responsible for monitoring the environment and intervening when automation degrades or fails \cite{on2021taxonomy}. The majority of design-focused studies still center on TOR content, modality, and timing \cite{du2021designing,lee2023investigating,huang2024enhancing}. While the \textit{silent failures} pose an equally, if not more, critical safety risk \cite{kanaan2024automation}. As highlighted in recent reviews and taxonomies of driving automation failures, such failures can arise at different stages of the automated driving pipeline, including perception, decision-making, and control \cite{kanaan2024automation}. Real-world crashes (e.g., Tesla/Florida, 2016; Uber/Tempe, 2018) illustrate how perception-level failures can escalate when drivers remain unaware of imminent hazards \cite{ntsb2017,ntsb2019}. Analytical work has found that such failures may arise at the perception layer of ADS \cite{chen2023where} and that the failures in perception may occur even when downstream planning appears reasonable. Thus, in this work, we focus on a practically important subset of silent failures in SAE Level 2 systems, in which the ADS’s perceived and/or planned state of the world is misaligned with the actual hazard on the road and therefore no TOR is issued. Operationally, this manifests as hazards that are not correctly marked in the perception layer and/or as planned trajectories that conflict with the hazard, which are made visible to the driver through the PSAE interfaces.

Beyond incident analyses, recent human-factors experiments have begun to reveal the risks of silent failures. \citet{Louw2019TRF} showed that, following silent lane-keeping failures, drivers engaged in non-driving-related tasks had slower responses and exhibited more lane excursions. \citet{Piccinini2020HF} reported delayed braking and degraded takeover quality in silent failure conditions relative to manual driving. \citet{Mole2020PLOS} modeled takeover timing in silent failures and demonstrated how individual and situational factors systematically affect response latency. However, very few researchers have explored mitigation strategies to improve takeover performance in silent failures. For example, projecting planned trajectories on the windshield has been shown to improve takeover responses in silent failures \cite{jung2023projecting}, and AR-HUDs can assist drivers during malfunctions \cite{feierle2022augmented}. However, their limited types of PSAE and relatively simple hazardous scenarios are unable to comprehensively assess the support provided by different PSAE for driving safety in dynamic traffic environments when silent failures occur.

\subsection{Interpretability and Explanations for Automated Driving Systems}
Interpretability refers to an ADS’s ability to provide meaningful and understandable information about its status, maneuvers, and uncertainties \cite{doshi2017towards,omeiza2021explanations}. As ADS increasingly rely on complex AI, their internal reasoning often diverges from human expectations, challenging user understanding and adversely affecting users' trust \cite{miller2019explanation}. A wide range of explainable-AI approaches have been proposed to increase transparency in driving contexts, including attention/saliency visualizations \cite{chitta2021neat,cultrera2023explaining}, interpretable perception outputs such as object detection and semantic segmentation \cite{abukmeil2021towards,sun2020see}, planned trajectory visualizations \cite{ono2019improvement,colley2022effects}, and uncertainty representations that signal system confidence \cite{michelmore2020uncertainty,tai2019visual}.

Situation awareness (SA) is widely defined as “the perception of the elements in the environment within a volume of time and space, the comprehension of their meaning, and the projection of their status in the near future” \cite{endsley1995measurement}. In partially automated driving, HMI perception-oriented visualizations mainly support Level 1 SA (“what is there”), while explanations at the decision level relate more to how these perceptions are used for maneuvers (Level 3 SA, “what will happen next”). This view is consistent with classical hierarchical models of driving \cite{michon1985critical}, where maintaining sufficient tactical-level SA is a prerequisite for safe operational control during a takeover. Most deployed or evaluated explanations at \textit{decision-level} (e.g., 'what the car will do'), implicitly assumed that the perception is accurate. This assumption breaks down in \textit{silent failure} scenarios, where the core problem is a misperception of the environment rather than a poor decision \cite{cummings2014man}. Recent works have started to focus on \textit{perception-level} transparency: e.g., visualizing recognized elements in a scene or system predictions to improve drivers’ SA in TOR-initiated scenarios \cite{colley2021chi,colley2022effects,goodge2024can}. Yet, how prospective, perception-level information (what the system 'currently sees' and 'expects to happen next') affects drivers’ psychological readiness, physiological preparation, and ultimate takeover performance under silent failures remains underexplored.

Beyond the content in the displays, researchers have also explored how explanations can be delivered to drivers. HUD-based explanations, such as AR overlays of hazards or trajectories, can improve takeover quality \cite{feierle2022augmented}. Multimodal approaches, combining visual with auditory explanations, can further reduce cognitive load \cite{du2021designing,lee2023investigating}. However, while prior works have explored head–up display for providing planned trajectories or perceptual cues, they have largely focused on TOR scenarios or have not systematically investigated the combined effects of both perception and planned maneuver information in silent failure conditions. 

\begin{figure*}
\begin{center}
\includegraphics[scale=0.65]{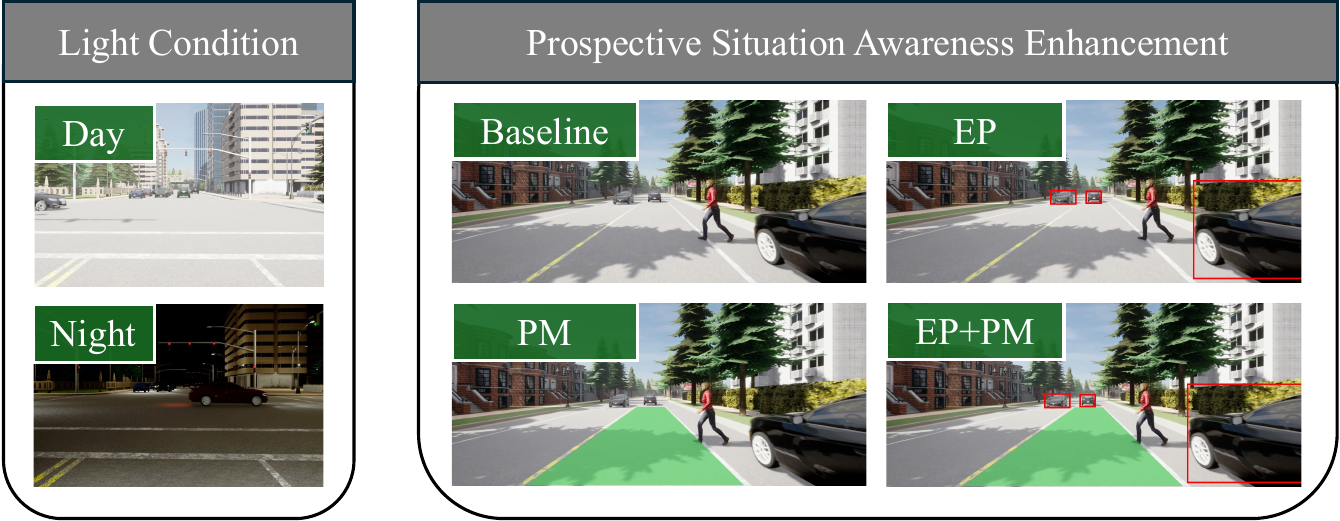}\\
\end{center}
\vspace{-4mm}
\caption{Illustration of the two between-subject conditions, including four PSAEs and two light conditions presented to the driver in the simulator.}\label{scenario}
\vspace{-4mm}
\end{figure*}

\subsection{This Study}
To address the above-mentioned gaps, our study aims to examine how perception-level, forward-looking information, both alone and combined with environmental factors, can affect drivers' internal states and safety-critical behavior in L2 driving. Specifically, our work simulated a dynamic AR-HUD, offering continuous perception- and projection-level information to support drivers in silent failure scenarios, as user studies indicate that such explanations can bolster trust, SA, and takeover quality—especially when prompted by TORs \cite{kim2023and,lee2023investigating,huang2024enhancing,du2021designing}. Further, given that lighting conditions, hazard visibility, and task workload have been shown to affect trust, attention allocation, and takeover performance \cite{de2014effects,wang2025exploring,wang2024revisiting}, we further evaluated how drivers’ psychological states (i.e., self-reported SA, perceived safety, and trust), their EEG- and EMG-based physiological measures, and the manipulated environmental factors (day/night and visible/invisible hazards) are dynamically coupled when different PSAE information is provided on an AR-HUD.

\section{Methodology}

\subsection{Participants}
We recruited 48 participants (24 male, 24 female) from the local community via online postings and campus mailing lists. Eligible participants must have held a valid driving licence for at least two years, own a vehicle with ADS or self-report to be familiar with ADS features (such as Autopilot in Tesla or equivalent systems), and have driven at least 5,000 km in the past year. Drivers' age ranged from 23 to 42 ($M=27.54$, $SD=4.60$), with years of licensure from 2 to 19 ($M=6.52$, $SD=3.57$). This population represents active, non-novice urban drivers who are more likely to be early adopters of SAE Level-2 automation \cite{wang2024trust, huang2024exploring,nordhoff2022profiling}. Participants self-reported having normal or corrected-to-normal vision and no neurological or motor impairments. Each participant received 100 CNY for participating in this study. The study protocol was approved by the Human and Artefacts Research Ethics Committee at the Hong Kong University of Science and Technology (Guangzhou) (protocol: HSP-2023-0009).


\subsection{Prospective Situation Awareness Enhancement}
Our Prospective Situation Awareness Enhancement (PSAE) was designed with three guiding objectives: (1) \textit{Transparency of perception}, i.e., what ADS detects, so drivers can notice perceptual misses; (2) \textit{Action affordance}, which links perception to planned maneuvers so that drivers can quickly infer whether the system intends to avoid a detected object or not; (3) \textit{Low intrusion \& stability}, which provides continuous information that improves anticipation without producing attentional overload or distracting flicker \cite{colley2021chi,colley2022effects,feierle2022augmented}.

Subsequently, except for the baseline condition (no PSAE applied), we implemented three PSAE visualization strategies, each rendered as dynamic overlays on the forward view (see Figure \ref{scenario}):

\begin{itemize}
  \item \textbf{Environment Perception (EP).} ADS-detected objects are highlighted using color-coded bounding boxes and textual labels. Color mapping: \textit{yellow} for traffic infrastructure (e.g., lights/signs), \textit{blue} for pedestrians, and \textit{red} for vehicles/obstacles. The numeric confidence score is not presented to avoid excessive cognitive load. \cite{ribeiro2016should,lundberg2017unified,kendall2017uncertainties,chitta2021neat}.
  \item \textbf{Planned Maneuver (PM).} The ADS’s short-term planned trajectory is rendered as a semi-transparent green polyline extending from the ego-vehicle for a 3–5 s horizon, together with a shaded stopping envelope, indicating the current braking/steering safety margin. The trajectory rendering communicates intended lateral and longitudinal actions and visually shows potential future conflicts with overlaid objects \cite{ono2019improvement,jung2023projecting}.
  \item \textbf{Combined (EP+PM).} Both EP and PM are presented concurrently, enabling drivers to directly map detected elements to the system’s intended maneuver and thus quickly identify perception–action mismatches (e.g., a missing bounding box while the PM path intersects an undetected pedestrian).
\end{itemize}

Because the experiment studies \emph{silent failures} (perceptual misses without a TOR), we implemented two types of overlays per traffic scenario. Normal operation: overlays faithfully reflect the ADS detections and planned trajectory recorded during the Wizard-of-Oz run (bounding boxes present, PM avoids hazards). Silent failure: overlays simulate perception misses by omitting expected bounding boxes (EP miss) and/or showing a planned trajectory that intersects the hazard (PM mismatch). Importantly, in silent-failure trials, \emph{no} takeover request or alarm is presented; PSAE is the only feedforward information available to the driver. All visual elements were designed to maintain legibility and minimize clutter under both day and night conditions.

\begin{table}[ht]
\centering
\scriptsize
\caption{Categories of silent-failure scenarios in the study}
\label{tab:scenarios}

\setlength{\tabcolsep}{2pt}
\renewcommand{\arraystretch}{1.1}

\begin{tabularx}{\columnwidth}{c X m{0.26\columnwidth}}
\toprule
\textbf{Hazard visibility} & \textbf{Scenario description} & \textbf{Illustration} \\
\midrule
\multirow{3}{*}{\textbf{Invisible}} 
& Scenario 1: Left-turning at an unsignalized junction with the view obstructed by trees and fences and roadside objects (visible $\sim$50\,m), indicating possible cross-traffic.
& \includegraphics[width=\linewidth]{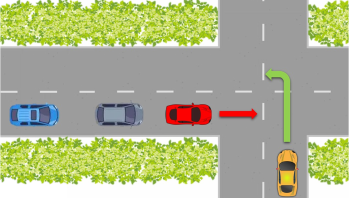} \\[0.8em]

& Scenario 5: Pedestrian stepping into the road from behind an advertising board on a curved road; the board (visible $\sim$60\,m) signals possible hidden pedestrians.
& \includegraphics[width=\linewidth]{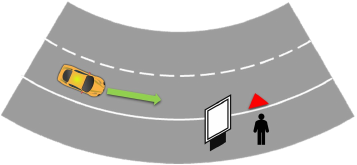} \\[0.8em]

& Scenario 6: Pedestrian emerging from behind a parked vehicle when ego car goes straight; the parked car (visible $\sim$70\,m) serves as the cue.
& \includegraphics[width=\linewidth]{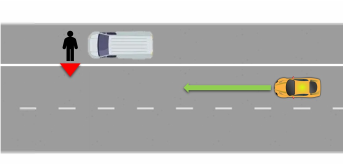} \\[0.8em]

\midrule
\multirow{3}{*}{\textbf{Visible}} 
& Scenario 2: Pedestrian suddenly crossing from the right at a signalized intersection against the red light.
& \includegraphics[width=\linewidth]{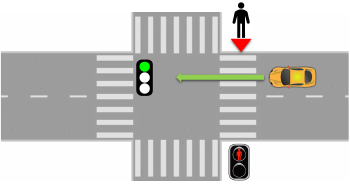} \\[0.8em]

& Scenario 3: Parked car in the horizon on the right pulling out and merging abruptly into the ego lane.
& \includegraphics[width=\linewidth]{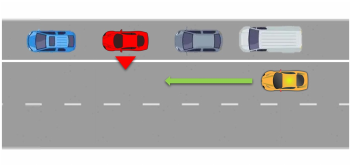} \\[0.8em]

& Scenario 4: Encountering a road construction zone while driving straight ahead.
& \includegraphics[width=\linewidth]{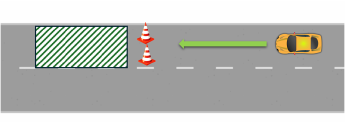} \\
\bottomrule
\end{tabularx}

\medskip
\footnotesize\textbf{Note:} The yellow car indicates the ego vehicle with green arrows showing its intended path. Hazards are represented in red: continuous arrows for ongoing motion and triangles for the onset of movement.
\end{table}

\subsection{Experimental Design and Procedures}
The study employed a mixed 4×2×2 factorial design with one within-subject factor and two between-subject factors. The between-subject factors were PSAE type (Baseline, EP, PM, and EP+PM) and lighting condition (day vs. night). The within-subject factor was hazard visibility (visible vs. invisible hazards). The six silent-failure events consisted of three scenarios with visible hazards and three scenarios with invisible hazards (Table \ref{tab:scenarios}). Each participant experienced all six events exactly once, under a single PSAE type and lighting condition, such that every driver encountered three visible-hazard and three invisible-hazard failures in one continuous loop drive. 

\begin{figure}
\begin{center}
\includegraphics[scale=0.43]{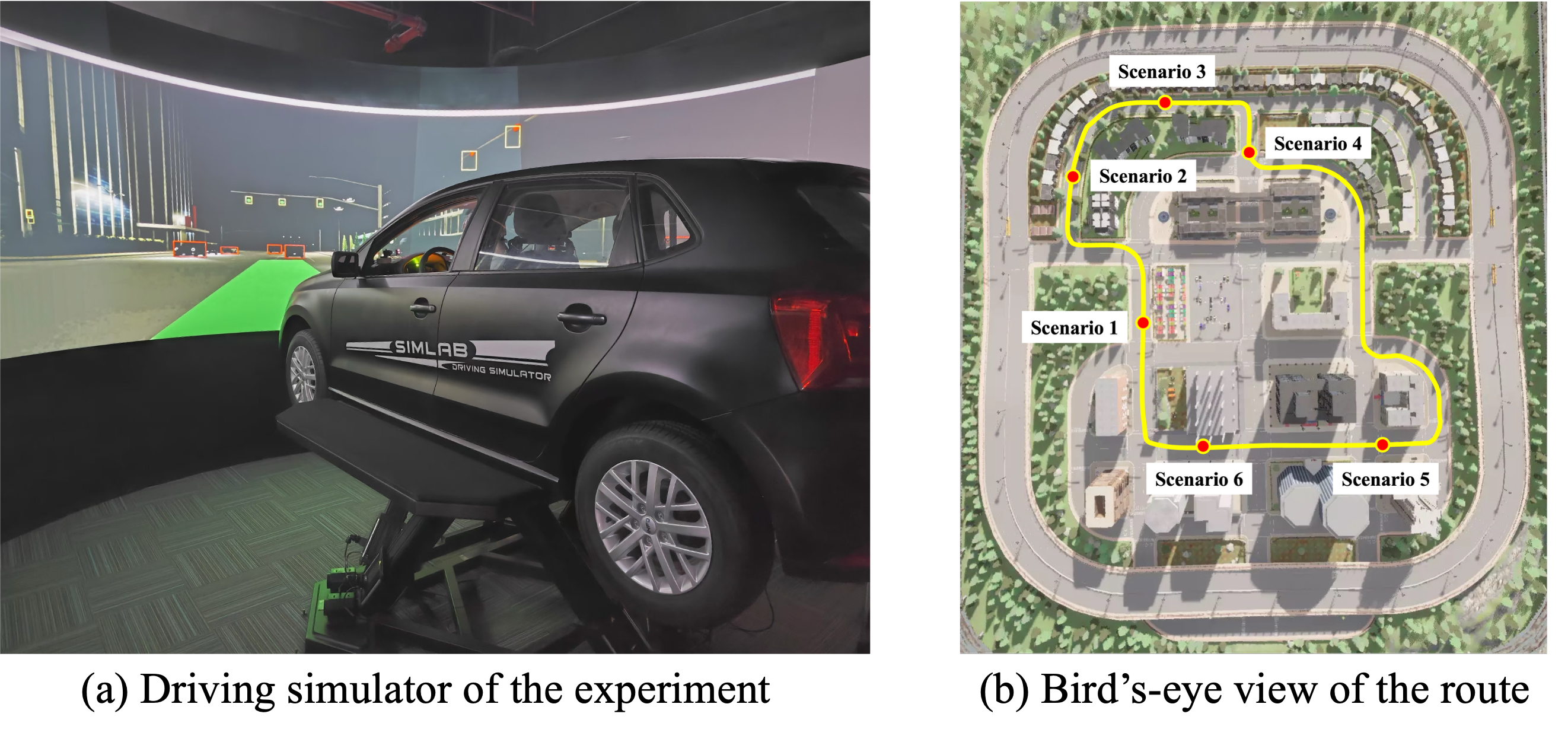}\\
\end{center}
\vspace{-4mm}
\caption{The demonstration of (a) six-degree-of-freedom driving simulator and (b) experimental route setup.}\label{apparatus}
\vspace{-4mm}
\end{figure}

\textit{Hazard visibility}. Following \cite{yan2024ch}, visible hazards are directly observable in the driver’s forward field of view when they enter the scene (e.g., a pedestrian already in view crossing against a red light, a clearly marked construction zone, or a parked vehicle that is continuously visible before pulling out). In contrast, invisible hazards are initially occluded by other objects and must be inferred from contextual cues (e.g., a pedestrian emerging from behind an advertising board or a parked vehicle, or cross-traffic hidden behind roadside structures). To give participants enough time to develop situation awareness, all contextual cues (for invisible hazards) or hazard visibility horizons (for visible hazards) are from 8 to 15 seconds, exceeding the recommended minimum of 5 seconds \cite{mok2015timing}. The six events were placed at fixed locations along a closed urban route in CARLA Town 5 (Fig.~\ref{apparatus}(b)) \cite{dosovitskiy2017carla}. Scenario order was counterbalanced within each PSAE × lighting group using a Latin-square scheme by varying the starting point on the loop, such that each scenario appeared equally often in each ordinal position across the six drivers in a group.

\textit{Lighting}. We implemented two ambient lighting conditions in CARLA, corresponding to a clear daytime and a nighttime urban environment. In the day condition, the simulation time-of-day was set to midday with high ambient illumination and no artificial street lighting. In the night condition, the time-of-day was set to evening with low ambient illumination, active street lights, and ego-vehicle headlights, while keeping road geometry, traffic, and hazard timings identical to the day scenarios. These settings ensured that only the lighting conditions, rather than the underlying traffic dynamics, differed between day and night.

Before driving, participants were invited based on the screening questionnaire (including demographics and driving history information) and first completed informed consent. To mitigate the potential influence of different ADS usage experience on drivers’ responses to silent failures, they then received standardized instruction and completed a practice/demo drive containing one practice silent-failure event (not used in the formal experiment) to familiarize them with the manual driving of the vehicle, how to take over the control of the vehicle, and the assigned PSAEs. During the experiment, takeover was detected when a participant initiated steering beyond $\pm10^{\circ}$ or depressed the brake pedal beyond 10\% travel, consistent with prior studies \cite{li2021adaptive}. 

Then, the formal experiment started, which contained six silent-failure events. Whenever a takeover was detected throughout the drive, the simulator paused. Takeovers occurring outside the designated event windows (around 10 seconds from first visibility of the hazard/cue object to collision) were classified as error takeovers. If it happened, the drive resumed from the paused point. Successful takeovers caused the remainder of that failure segment to be skipped to simulate successful avoidance. We further extracted the \textit{Takeover Lead Time} and \textit{Takeover Success} for further analysis.

After each takeover or scenario end (if no takeover was detected in a scenario, which lasted around 5 minutes), participants completed a 10-item Situation Awareness questionnaire for assessing situation awareness \cite{taylor2017situational}, a TiAD trust scale \cite{manchon2022initial}, and a Perceived Safety scale adopted from \cite{cao2021development}.

\subsection{Apparatus}
Experiments were run on a six-degree-of-freedom motion-capable driving simulator (WIVW GmbH) with a realistic vehicle cabin (force-feedback steering, pedals). Visuals were presented via projection over a 210° horizontal field of view at a combined resolution of 5760×1200 pixels to create an immersive urban environment (Fig.~\ref{apparatus}). To ensure repeatability and precise control over silent failures, we adopted a Wizard-of-Oz approach: experimenters manually drove and recorded scenario runs in CARLA to produce deterministic routes and failures during autonomous driving. A custom Python playback application rendered the recorded frames of the vehicle front camera, applied PSAE overlays frame-by-frame, and synchronized playback with participants' steering and braking inputs; steering angle and pedal signals were logged over Ethernet for offline alignment.

Physiological measures comprised continuous EEG recorded with a Neuroelectrics Enobio 32-channel dry sensors\footnote{https://www.neuroelectrics.com/products/research/enobio/enobio-32} at 500 Hz, and surface EMG recorded at 1000 Hz through PhysioLAB platform\footnote{https://www.infoinstruments.cn/product/physiolab/} from two sites: the dominant-hand forearm (approximately the flexor region, to index steering-related activation) and the right medial gastrocnemius (to index braking/leg activation) \cite{satti2021microneedle,tjolleng2022analysis}. All data streams (simulator telemetry, AR-HUD rendering timestamps, EEG, EMG, and behavioral logs) were synchronized via millisecond timestamps to enable precise multimodal analyses.

\begin{figure*}
\begin{center}
\includegraphics[scale=0.5]{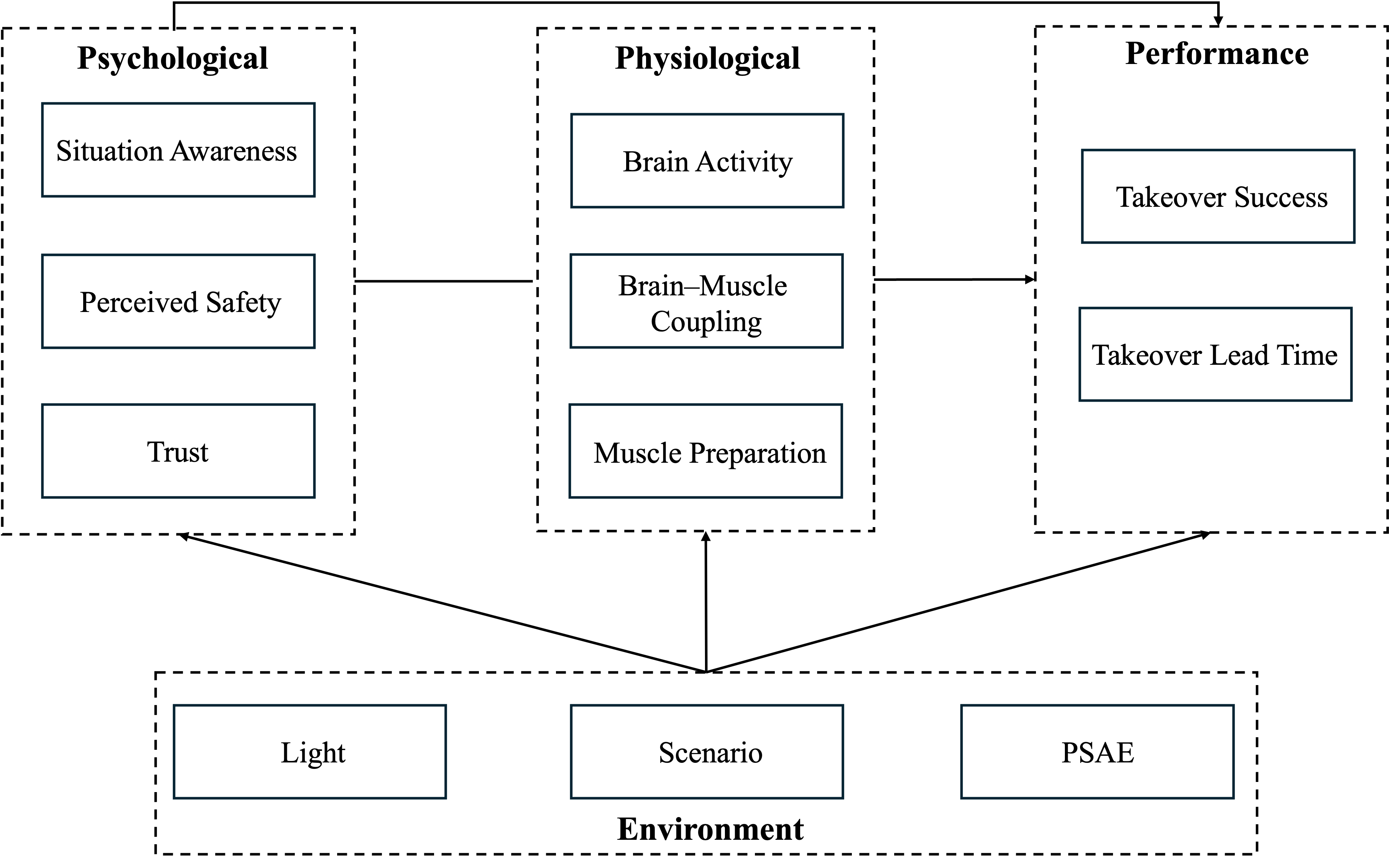}\\
\end{center}
\vspace{-4mm}
\caption{The final theoretical framework of the model.}\label{framework}
\vspace{-4mm}
\end{figure*}

\subsection{Data Processing and Variable Extraction}
\paragraph{EEG} The preprocess of EEG signal followed a standardized MNE pipeline \cite{GramfortEtAl2013a} to ensure high signal quality and reproducibility: band-pass filtering between 1–30 Hz was applied to remove slow drifts and high-frequency noise, common average re-referencing, and Independent Components Analysis (ICA) was used to isolate and remove components associated with eye blinks and ocular movements \cite{delorme2004eeglab}. Channels with excessive noise (variance > 3 standard deviations from the mean) were interpolated using spherical spline interpolation \cite{perrin1989spherical}. To quantify neural dynamics related to experimental events, we computed the relative ERSP change ratio (\textit{Power Ratio}), which measures changes in rhythmic brain activity power relative to the pre-event baseline period.

Specifically, for each 10-second event, we extracted a 4-second pre-event baseline from the raw EEG time series. The baseline series was converted to a time-frequency representation via Morlet wavelets, which provides an optimal trade-off between temporal and frequency resolution for analyzing cognitive processes~\cite{cohen2014analyzing}. For each scene's event and baseline power, we computed the median power per channel and per frequency across time. The resulting power at each time point ($P_{\text{event}}(t)$) at every event window was normalized against the median power from a 4-second pre-event baseline period ($P_{\text{baseline}}$) using the formula:

\begin{equation}
\text{ERSP}(t) = \frac{P_{\text{event}}(t) - P_{\text{baseline}}}{P_{\text{baseline}}}.
\label{eq:ersp}
\end{equation}

We summarized ERSP within canonical frequency bands (i.e., $\theta$ (4–8 Hz), $\alpha$ (8–12 Hz), and $\beta$ (15–30 Hz)) and averaged these across three large-scale Region-of-Interests (ROIs) (i.e., Frontal, Sensorimotor, Parietal) to capture dynamics related to attention, motor planning, and sensory integration \cite{maksimenko2025video}. From each ROI–band time–frequency matrix, we extracted two features: (1) the median \textit{Power Ratio} across the event windows time, and (2) \textit{Activation Time}, the latencies of the peak responses (for Beta the time of maximal synchronization; for Theta/Alpha the time of maximal desynchronization \cite{pfurtscheller1999event}), which together characterize both magnitude and temporal dynamics of neural responses.

\begin{table}[htbp]
\centering
\scriptsize
\caption{Summary of dependent variables across different layers.}
\label{variables}
\renewcommand{\arraystretch}{1.2} 
\setlength{\tabcolsep}{2pt} 
\begin{tabularx}{\columnwidth}{lll}
\toprule
\textbf{Layer} & \textbf{Variable [unit]} & \textbf{Distribution} \\
\midrule
\multirow{3}{*}{Psychological Perception} 
& Situation Awareness [-] & Mean: 5.2 (SD: 0.9, min: 2.3, max: 7) \\
& Perceived Safety [-] & Mean: 4.1 (SD: 1.0, min: 1.0, max: 6) \\
& Trust [-] & Mean: 45.7 (SD: 27.1, min: 0, max: 91) \\
\midrule
\multirow{6}{*}{Cognitive Activity} 
& $S_\beta$ Activation Time [ms] & Mean: 3613 (SD: 3519, min: 2, max: 10000) \\
& $F_\theta$ Activation Time [ms] & Mean: 5510 (SD: 2633, min: 2, max: 10000) \\
& $P_\alpha$ Activation Time [ms] & Mean: 5192 (SD: 2669, min: 112, max: 9828) \\
& $S_\beta$ Power Ratio [-] & Mean: 0.3 (SD: 2.2, min: -0.9, max: 5.5) \\
& $F_\theta$ Power Ratio [-] & Mean: 1.05 (SD: 10.4, min: -1, max: 23.7) \\
& $P_\alpha$ Power Ratio [-] & Mean: 0.3 (SD: 2.3, min: -1, max: 29.3) \\
\midrule
\multirow{2}{*}{Muscle Preparation} 
& Preparation [-] & \makecell[l]{No preparation (n=28, 12.8\%) \\ Preparation exist (n=190, 87.2\%)} \\
& Preparation Time [ms] & Mean: 3347 (SD: 3885, min: 0, max: 9999) \\
\midrule
\multirow{3}{*}{Brain--Muscle Coupling} 
& $S_\beta$ Time Lag [ms] & Mean: 266.3 (SD: 4481, min: -9977, max: 10000) \\
& $F_\theta$ Time Lag [ms] & Mean: 2163 (SD: 4687, min: -9378, max: 10000) \\
& $P_\alpha$ Time Lag [ms] & Mean: 1845 (SD: 4588, min: -9862, max: 9120) \\
\midrule
\multirow{2}{*}{Performance} 
& Takeover Lead Time [ms] & Mean: 1106 (SD: 982, min: 0, max: 5800) \\
& Takeover Success [-] & \makecell[l]{No takeover (n=32, 14.5\%) $\sim$ 0 \\ Success takeover (n=186, 85.5\%) $\sim$ 1} \\
\bottomrule
\end{tabularx}
\textnormal{Notes: $S_\beta$: Sensorimotor Beta; $F_\theta$: Frontal Theta; $P_\alpha$: Parietal Alpha; SD: standard deviation}
\end{table}

\paragraph{EMG} To quantify participants' motor preparation, we processed surface EMG signals recorded from the hand and leg. The raw signals first underwent preprocessing, which included a zero-phase fifth-order Butterworth band-pass filter to isolate the primary EMG frequency content~\cite{de1997use}, followed by full-wave rectification (i.e., taking the absolute value) and normalization to prepare the signal for amplitude analysis~\cite{phinyomark2018feature}. We then identified muscle activation onset by detecting the first sample in the rectified signal that exceeded a predefined activation threshold 0.1 (i.e., 10\% of the normalized EMG amplitude \cite{carvalho2023review}) for EMG-based onset detection~\cite{hodges1996comparison}. This onset time, calculated in milliseconds relative to the end of the trial window, served as our primary measure of motor preparation latency. Furthermore, based on the binary activation status of the hand and leg channels, we classified each trial into preparation exists and no preparation (\textit{Preparation}), and overall \textit{Preparation Time} was defined as the earliest muscle activation onset across the measured limbs.

\paragraph{Brain–Muscle Coupling} 
To characterize the temporal coordination between cortical processes and peripheral muscle execution, we computed the latency difference (\textit{Time Lag}) between the peak of the ERSP in the EEG and the onset of muscle activation in EMG:

\begin{equation}
Time\ Lag = t_{\text{EEG\_peak}} - t_{\text{EMG\_onset}}.
\end{equation}

While the onset of motor-related potentials (e.g., the Readiness Potential) of EEG offers a more direct precursor to movement, the ERSP peak was chosen for two reasons. First, it represents the point of maximal cortical disinhibition and engagement in motor planning for this reactive task, capturing the climax of the decision-to-act process \cite{pfurtscheller1999event}. Second, peak detection is more robust to the signal-to-noise limitations inherent in dry-EEG recordings compared to threshold-based onset detection. Therefore, our latency measure characterizes the timing between peak cortical activity and peripheral execution. A summary of all extracted variables and descriptive statistics is provided in Table~\ref{variables}.

\begin{table*}[h]\centering
\caption{Summary of statistical results.}
\label{statistic}
\small
\begin{tabular}{c|c|c|c|c|c}
\toprule[1pt]
\textbf{Dependent Variable (DV)} & \textbf{Independent Variable (IV)} & \textbf{F-value/$\chi^2$-value} & \textbf{Estimate (95\% CI)} & \textbf{\textit{p}} & \textbf{Fit indicator} \\
\hline
\multirow{2}{*}{Situation Awareness} 
  & Light & F(1, 40) = 4.69 & -- & .03$^{*}$ 
  & \multirow{2}{*}{\begin{tabular}[c]{@{}c@{}}Adj $R^2$ = 0.08\\$p$ = .001$^{*}$\end{tabular}} \\
  & PSAE  & F(3, 40) = 5.11 & -- & .002$^{*}$ & \\
\hline
\multirow{3}{*}{Perceived Safety} 
  & PSAE     & F(3, 42) = 1.85  & n.s. & .1 
  & \multirow{3}{*}{\begin{tabular}[c]{@{}c@{}}Adj $R^2$ = 0.15\\$p$ < .0001$^{*}$\end{tabular}} \\
  & Light    & F(1, 40) = 1.00  & n.s. & .2 & \\
  & Scenario & F(1, 40) = 30.81 & --   & $<$.0001$^{*}$ & \\
\hline
\multirow{3}{*}{Trust} 
  & PSAE     & F(3, 40) = 6.43  & -- & .0003$^{*}$ 
  & \multirow{3}{*}{\begin{tabular}[c]{@{}c@{}}Adj $R^2$ = 0.12\\$p$ < .0001$^{*}$\end{tabular}} \\
  & Scenario & F(1, 42) = 5.57  & -- & .02$^{*}$       & \\
  & Light    & F(3, 40) = 4.61  & -- & .03$^{*}$       & \\
\hline
\multirow{5}{*}{Preparation Time} 
  & PSAE                      & F(3, 42) = 2.69  & --   & .048$^{*}$ 
  & \multirow{5}{*}{\begin{tabular}[c]{@{}c@{}}Adj $R^2$ = 0.07\\$p$ = .03$^{*}$\end{tabular}} \\
  & Light                     & F(1, 40) = 1.01  & n.s. & .3         & \\
  & Scenario                  & F(1, 40) = 0.44  & n.s. & .5         & \\
  & Trust                     & F(1, 238) = 0.72 & n.s. & .4         & \\
  & Situation Awareness*PSAE  & F(3, 238) = 1.86 & n.s. & .1         & \\
\hline
Preparation 
  & Light & $\chi^2$(1) = 6.64 & -- & .01$^{*}$ 
  & \begin{tabular}[c]{@{}c@{}}pseudo-$R^2$ = 0.09\\$p$ = .001$^{*}$\end{tabular} \\
\hline
\multirow{4}{*}{$F_\theta$ Activation Time} 
  & PSAE                     & F(3, 40) = 3.36  & --   & .02$^{*}$ 
  & \multirow{4}{*}{\begin{tabular}[c]{@{}c@{}}Adj $R^2$ = 0.11\\$p$ = .01$^{*}$\end{tabular}} \\
  & Scenario                 & F(1, 40) = 4.08  & --   & .045$^{*}$ & \\
  & Situation Awareness      & F(1, 238) = 2.72 & n.s. & .053       & \\
  & Situation Awareness*PSAE & F(3, 238) = 2.48 & --   & .0498$^{*}$& \\
\hline
\multirow{4}{*}{$P_\alpha$ Activation Time} 
  & Situation Awareness      & F(1, 239) = 4.41 & 431.42 [25.17, 837.66] & .04$^{*}$ 
  & \multirow{4}{*}{\begin{tabular}[c]{@{}c@{}}Adj $R^2$ = 0.04\\$p$ = .02$^{*}$\end{tabular}} \\
  & Perceived Safety         & F(1, 239) = 0.39 & n.s. & .5         & \\
  & PSAE                     & F(3, 42) = 1.07  & n.s. & .4         & \\
  & Situation Awareness*PSAE & F(3, 239) = 0.83 & n.s. & .5         & \\
\hline
\multirow{4}{*}{$F_\theta$ Power Ratio} 
  & PSAE                     & F(3, 42) = 4.12  & --   & .001$^{*}$ 
  & \multirow{4}{*}{\begin{tabular}[c]{@{}c@{}}Adj $R^2$ = 0.07\\$p$ = .045$^{*}$\end{tabular}} \\
  & Trust                    & F(1, 238) = 1.83 & n.s. & .2         & \\
  & Situation Awareness      & F(1, 238) = 3.91 & n.s. & .0501      & \\
  & Situation Awareness*PSAE & F(3, 238) = 3.21 & --   & .03$^{*}$  & \\
\hline
\multirow{3}{*}{$P_\alpha$ Power Ratio} 
  & PSAE                & F(3, 42) = 2.01  & n.s. & .2                         
  & \multirow{3}{*}{\begin{tabular}[c]{@{}c@{}}Adj $R^2$ = 0.05\\$p$ = .03$^{*}$\end{tabular}} \\
  & Situation Awareness & F(1, 238) = 7.31 & -0.48 [-0.83, -0.13] & .008$^{*}$ & \\
  & Trust               & F(1, 238) = 4.03 & -0.01 [-0.02, -0.0002] & .047$^{*}$ & \\
\hline
$P_\alpha$ Time Lag 
  & Situation Awareness & F(1, 239) = 4.98 & 907.03 [103.75, 1710.31] & .03$^{*}$ 
  & \begin{tabular}[c]{@{}c@{}}Adj $R^2$ = 0.03\\$p$ = .02$^{*}$\end{tabular} \\
\hline
\multirow{9}{*}{Takeover Success} 
  & $P_\alpha$ Activation Time & $\chi^2$(1) = 4.93 & 0.0002 [0.0, 0.0004] & .03$^{*}$ 
  & \multirow{9}{*}{\begin{tabular}[c]{@{}c@{}}pseudo-$R^2$ = 0.48\\$p$ < .0001$^{*}$\end{tabular}} \\
  & $P_\alpha$ Time Lag        & $\chi^2$(1) = 2.19 & n.s. & .1                & \\
  & Light                      & $\chi^2$(1) = 29.00 & --   & $<$.0001$^{*}$    & \\
  & PSAE                       & $\chi^2$(3) = 4.88  & n.s. & .2                & \\
  & Scenario                   & $\chi^2$(1) = 19.43 & --   & $<$.0001$^{*}$    & \\
  & Situation Awareness        & $\chi^2$(1) = 8.32  & 1.64 [0.47, 2.82] & .004$^{*}$ & \\
  & Perceived Safety           & $\chi^2$(1) = 8.33  & 0.96 [0.31, 1.61] & .004$^{*}$ & \\
  & Situation Awareness*PSAE   & $\chi^2$(3) = 8.93  & --   & .03$^{*}$         & \\
\hline
\multirow{3}{*}{Takeover Lead Time} 
  & Light               & F(1, 40) = 15.91 & -- & .0003$^{*}$ 
  & \multirow{3}{*}{\begin{tabular}[c]{@{}c@{}}Adj $R^2$ = 0.36\\$p$ < .0001$^{*}$\end{tabular}} \\
  & Scenario            & F(1, 40) = 8.08  & -- & $<$.0001$^{*}$ & \\
  & Situation Awareness & F(1, 235) = 4.85 & 157.11 [16.03, 298.19] & .03$^{*}$ & \\
\bottomrule[1pt]
\end{tabular}
\textnormal{\\Notes: In this table and the following tables, $^{*}$ marks significant results (\textit{p}<.05). In the Estimate (95\% CI) column, “n.s.” means nonsignificant (\textit{p} > 0.05), and “--” means the dependent variable is a nominal variable so that there are no Estimate values. The “Fit indicator” column reports, for each dependent variable, the adjusted $R^2$ (for linear mixed-effects models) or a McFadden-type pseudo-$R^2$ (for generalized linear models), and $p$-value when comparing the model to the null model without predictors.}
\end{table*}

\subsection{Statistical Analysis}

Based on previous theory frameworks \cite{michon1985critical, endsley1995measurement} and previous empirical studies \cite{wang2026modeling,colley2023effects,wang2024revisiting,du2020predicting,wang2025exploring}, our analytical approach is guided by a theoretical framework that decomposes the driver's response into a multi-layered causal sequence (see Figure~\ref{framework}). Our analytical approach was designed to systematically test the relationships within our multi-layered theoretical framework, which posits that the \textit{Environment} layer influences \textit{Psychological} layer. Then \textit{Psychological} layer is further associated with \textit{Physiological} layer, and they jointly or independently affect the final takeover \textit{Performance} layer. 

The analysis proceeded in two main stages: preliminary regression analyses, followed by a formal path analysis to test for mediation. First, to establish the foundational relationships between the layers, we conducted a series of regression analyses using \texttt{SAS OnDemand for Academics}. The model type was chosen based on the outcome variable: Generalized Linear Models (via \texttt{PROC GENMOD}) for binary outcomes (i.e., takeover success), and Linear Mixed-effects Models (via \texttt{PROC MIXED}) for continuous variables such as takeover time. To properly handle the non-independence of repeated measures, all mixed models incorporated a random intercept for each participant, thereby accounting for individual differences. Initial models included the main effects of our experimental conditions and variables from the preceding layers (see Figure~\ref{framework}). 

At the same time, we are interested in the moderating effects of PSAE in this process. Therefore, when constructing the models for the variables in the \textit{Physiological} and \textit{Performance} layers, we incorporated two-way interaction between SA and PSAEs as independent variables. A backward stepwise selection procedure was then used to arrive at the most parsimonious model, with multicollinearity confirmed to be absent via the Variance Inflation Factor (VIF). Following significant main or interaction effects (\textit{p}<.05), Tukey-Kramer post-hoc tests \cite{kramerss1956extension} were performed. Note that, before fitting the models, we performed power checks based on our experiment design. Following \cite{cohen2013statistical}, we considered medium (f=0.25) and large (f=0.40) effects. Results suggested that our sample size affords high power to detect large main effects of PSAE and lighting (power>0.85 for f=0.40), but only moderate power for medium effects (power=0.43 for PSAE and power=0.61 for lighting condition) and limited power for interactions across multiple experimental factors. To avoid underpowered and unstable estimates under these constraints, we therefore (a) focused on main effects in the mixed-effects and generalized linear models, and (b) restricted interaction terms to a small set of two-way interactions between experimental factors that showed significant main effects and continuous psychological or physiological measures (e.g., SA × PSAE), without specifying interactions among experimental factors.

Finally, to further explore how the PSAE affects the takeover performance, we conducted a path analysis using Structural Equation Modeling (SEM) via the \texttt{lavaan} package in R \cite{rosseel2012lavaan}. To account for repeated measurements, we adopted the Mixed-Effects Model (MLM) in SEM to account for individual and experimental sequence effects \cite{wang2025exploring}. This approach avoids the violation of the SEM assumptions on data independence. To compare the effects of each PSAE condition against the baseline, we created dummy variables for the \textit{PSAE} factor. For the continuous variable, we used a robust maximum likelihood estimator (\texttt{MLR}) and tested the significance of the indirect effect using bias-corrected confidence intervals from 5,000 bootstrap resamples. For the binary variable, we declared it as an ordered factor and used the robust weighted least squares estimator (\texttt{WLSMV}), which is appropriate for categorical outcomes. For both models, our primary goal was to quantify and test the significance of the specific indirect effect for each PSAE condition, thereby revealing the underlying mechanism through which the interface influenced driver performance.

\begin{table*}[h]\centering
\caption{Significant Post-hoc Results for Discrete Independent Variables.}
\label{posthoc}
\small
\begin{tabular}{c|c|c|c|c|c|c}
\toprule[1pt]
\textbf{DV} & \textbf{IV} & \textbf{IV Level} & \textbf{IV Level compared to} & \textbf{Estimation (95\% CI)} & \textbf{t value/z score} & \textbf{\textit{p}} \\
\hline
\multirow{3}{*}{Situation Awareness} & Light & Night & Day & -0.25 [-0.48, -0.02] & t(40) = -2.17 & .03$^{*}$ \\
& PSAE & EP & EP+PM & 0.45 [0.04, 0.87] & t(42) = 2.86 & .02$^{*}$ \\
& & EP & Baseline & 0.61 [0.17, 1.04] & t(42) = -2.17 & .002$^{*}$ \\
\hline
Perceived Safety & Scenario & Visible & Invisible & 0.51 [0.33, 0.70] & t(40) = 5.55 & <.0001$^{*}$ \\
\hline
\multirow{4}{*}{Trust} & Light & Night & Day & -7.57 [-14.53, -0.62] & t(40) = -2.15 & .03$^{*}$ \\
& PSAE & PM & Baseline & 15.52 [2.35, 28.69] & t(42) = 3.05 & .01$^{*}$ \\
& & EP+PM & Baseline & 20.76 [8.01, 33.50] & t(42) = 4.22 & <.0001$^{*}$ \\
& Scenario & Visible & Invisible & 8.23 [1.36, 15.12] & t(40) = 2.36 & .02$^{*}$ \\
\hline
Preparation & Light & Night & Day & 1.42 [0.34, 2.50] & z = 2.58 & .01$^{*}$ \\
\hline
$F_\theta$ Activation Time & Scenario & Visible & Invisible & 765.35 [16.16, 1514.54] & t(20) = 2.02 & .045$^{*}$ \\
\hline
\multirow{2}{*}{Takeover Success} 
& Light & Night & Day & -2.70 [-3.69, -1.72] & z = -5.39 & <.0001$^{*}$ \\
& Scenario & Visible & Invisible & 2.23 [1.24, 3.22] & z = 4.41 & <.0001$^{*}$ \\
\hline
\multirow{2}{*}{Takeover Lead Time} & Light & Night & Day & -587.36 [-885.44, -289.28] & z = -3.99 & .0003$^{*}$ \\
& Scenario & Visible & Invisible & 1056.78 [828.42, 1285.14] & z = 9.39 & <.0001$^{*}$ \\
\bottomrule[1pt]
\end{tabular}
\textnormal{\\Notes: Non-significant variables are not listed in this table.}
\end{table*}

\begin{figure*}
\begin{center}
\includegraphics[scale=0.20]{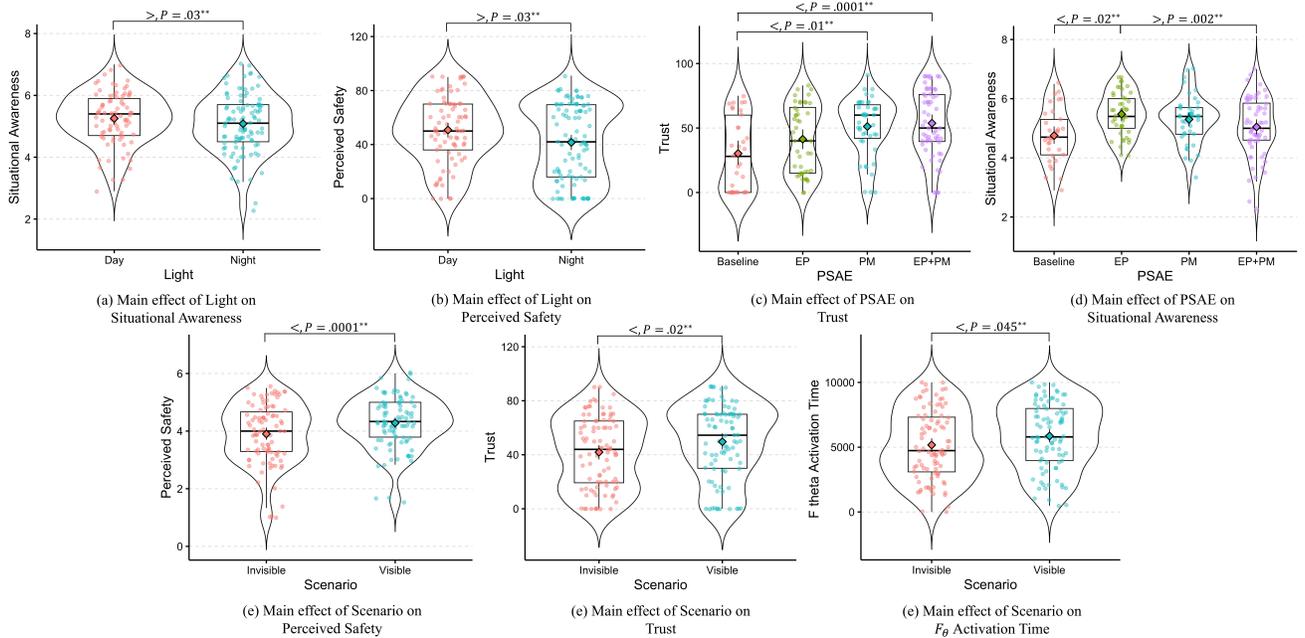}\\
\end{center}
\vspace{-4mm}
\caption{The main effects of experimental conditions on drivers' psychological and physiological states. (a) The effect of the time of day on drivers' situation awareness. (b) The effect of the time of the day on perceived safety. (c) The effect of PSAE information on trust. (d) The effect of PSAE information on situation awareness. (e) The effect of scenario type on perceived safety. (f) The effect of scenario type on trust. (g) The effect of scenario type on $F_\theta$ Activation Time.}\label{boxplot}
\vspace{-4mm}
\end{figure*}

\begin{figure*}
\begin{center}
\includegraphics[scale=0.07]{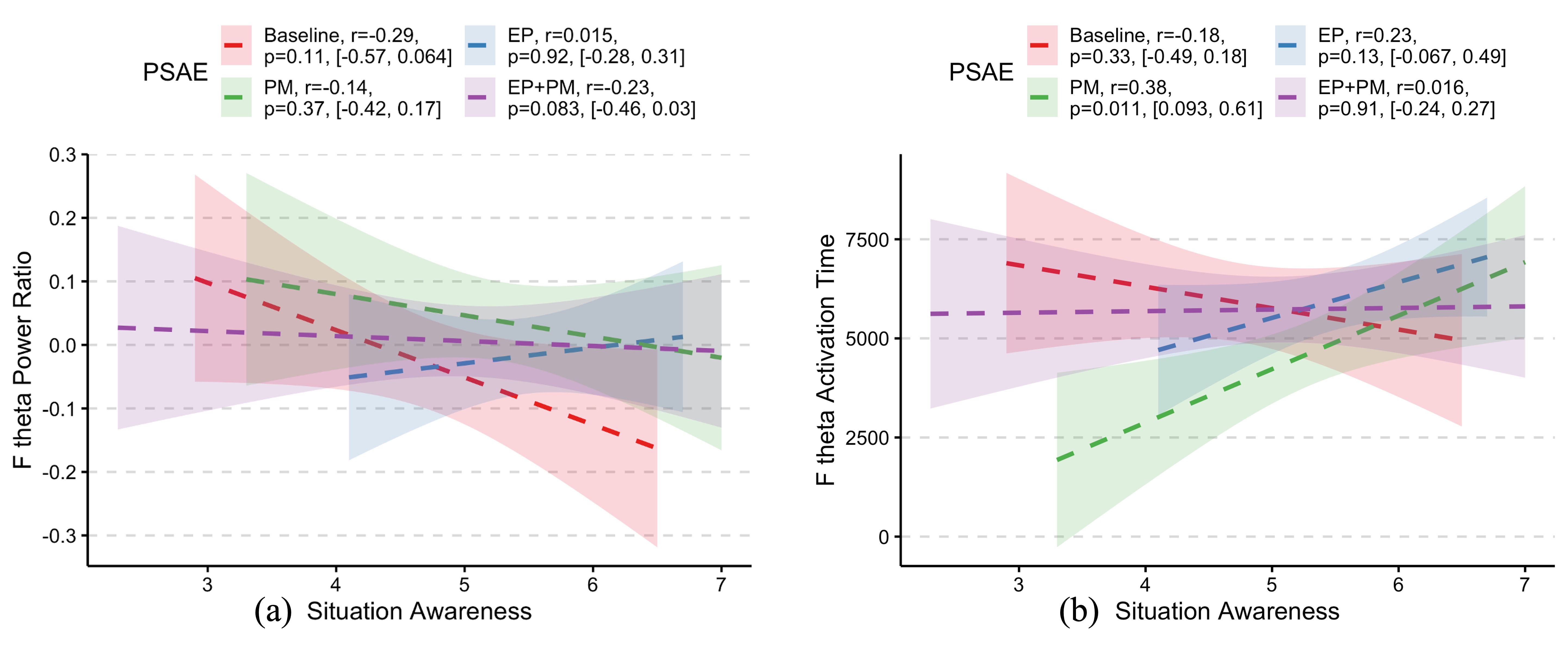}\\
\end{center}
\vspace{-4mm}
\caption{Interaction effects between the Situational Awareness and PSAE on (a) $F_\theta$ Power Ratio and (b) $F_\theta$ Activation Time. The samples of each PSAE group are fitted by a regression line with a
95\% confidence interval. Legends contain the results of correlation analysis based on Spearman’s correlation.}\label{scatplot}
\vspace{-4mm}
\end{figure*}

\section{Results}

\subsection{Regression Analysis}
This section presents the results from regression analyses, which were conducted as the first stage of our analytical approach to establish the foundational relationships within our theoretical framework (RQ1 and RQ2). The statistical summaries are presented in Table~\ref{statistic}, and the significant post-hoc comparisons are detailed in Table~\ref{posthoc}. In addition to the fixed-effect statistics, we also report overall model fit indicators. For continuous outcomes, we fit an equivalent fixed-effects linear model and report the resulting adjusted coefficient of determination (adjusted \(R^2\)) and \(p\)-value from the overall model \(F\)-test as descriptive indices of how much variance is accounted for by the fixed effects \cite{cohen2013applied}. For binomial outcomes, we report a McFadden-type pseudo-\(R^2\)  \cite{mcfadden1972conditional} and \(p\)-value in chi-square model test when comparing the current model versus intercept-only model. Visualizations of main effects and interaction effects are provided in Figure \ref{boxplot} and \ref{scatplot}, respectively. Note that, given the number of models and predictors considered, and the limited sample size in this study, the regression results in Table~\ref{statistic} should be viewed as exploratory. The key paths suggested by these exploratory regressions are subsequently evaluated in a more integrated manner in the SEM.

In response to RQ1, our regression results revealed that both the PSAE interface and lighting conditions were significant predictors of drivers' psychological states. Specifically, \textit{PSAE} had a significant main effect on both SA, which was measured using the SART scale, and Trust. The PM and EP+PM interfaces significantly increased driver \textit{Trust} compared to the Baseline condition. For \textit{Situation Awareness}, the results were more complex; the EP interface was associated with significantly higher \textit{Situation Awareness} compared to both the Baseline and the integrated EP+PM interface. The \textit{Light} condition also showed significant main effects on SA and Trust. Interestingly, drivers reported significantly lower \textit{Situation Awareness} during night-time drives compared to daytime, while simultaneously reporting lower \textit{Trust} in the system. Finally, the \textit{Scenario} type, as expected, was a powerful predictor for Perceived Safety and Trust. Environmental factors also influenced physiological readiness. The \textit{Light} condition had a significant effect on muscle \textit{Preparation}, with post-hoc tests showing a higher likelihood of preparation during night-time, corroborating the vigilance effect observed in the psychological data. The \textit{PSAE} interface also demonstrated a significant main effect on \textit{Preparation Time} and \textit{$F_\theta$ Power Ratio}. More importantly, we found significant interaction effects between \textit{PSAE} and \textit{SA} on several physiological variables, including $F_\theta$ Activation Time and $F_\theta$ Power Ratio, suggesting that the influence of the PSAE on drivers' neural activity is contingent upon their concurrent psychological state.

\begin{table*}[h]
\centering
\small
\caption{Path analysis results (a / b / c' / indirect).}
\label{path}
\begin{tabular}{c l l l l l}
\toprule
\textbf{DV} & \textbf{Path Type} & \textbf{Path} & \textbf{Estimation [95\% CI]} & \textbf{Std. $\beta$} & \textbf{$p$} \\
\midrule
\multirow{10}{*}{Takeover Success}
& a & EP $\to$ SA & 0.840 [0.369, 1.312] & 0.382 & <.0001$^{*}$\\
& a& PM $\to$ SA & 0.638 [0.196, 1.079] & 0.285 & .005$^{*}$\\
& a & EP+PM $\to$ SA & 0.409 [0.043, 0.775] & 0.201 & .03$^{*}$\\
& b  & SA $\to$ Takeover Success & 0.512 [0.330, 0.694] & 0.482 & <.0001$^{*}$\\
& c'   & EP $\to$ Takeover Success & 0.009 [-0.679, 0.696] & 0.004 & .98 \\
& c'   & PM $\to$ Takeover Success & -0.342 [-0.998, 0.313] & -0.144 & .3 \\
& c'   & EP+PM $\to$ Takeover Success & -0.018 [-0.636, 0.601] & -0.008 & .96 \\
& indirect (a$\times$b) & EP $\to$ SA $\to$ Takeover Success & 0.431 [0.173, 0.688] & 0.184 & <.0001$^{*}$\\
& indirect (a$\times$b) & PM $\to$ SA $\to$ Takeover Success & 0.327 [0.084, 0.570] & 0.137 & .008$^{*}$\\
& indirect (a$\times$b) & EP+PM $\to$ SA $\to$ Takeover Success & 0.209 [0.019, 0.400] & 0.097 & .03$^{*}$\\
\midrule
\multirow{10}{*}{Takeover Lead Time}
& a   & EP $\to$ SA & 0.805 [0.380, 1.231] & 0.354 & <.0001$^{*}$\\
& a   & PM $\to$ SA & 0.640 [0.193, 1.087] & 0.277 & .005$^{*}$\\
& a   & EP+PM $\to$ SA & 0.416 [-0.072, 0.903] & 0.198 & .095\\
& b  & SA $\to$ Takeover Lead Time & 0.211 [0.048, 0.374] & 0.186 & <.0001$^{*}$\\
& c'   & EP $\to$ Takeover Lead Time & 0.129 [-0.323, 0.581] & 0.050 & .6 \\
& c'   & PM $\to$ Takeover Lead Time & 0.141 [-0.330, 0.613] & 0.054 & .6 \\
& c'   & EP+PM $\to$ Takeover Lead Time & 0.361 [-0.067, 0.789] & 0.152 & .098\\
& indirect (a$\times$b) & EP $\to$ SA $\to$ Takeover Lead Time & 0.170 [0.018, 0.321] & 0.066 & .03$^{*}$\\
& indirect (a$\times$b) & PM $\to$ SA $\to$ Takeover Lead Time & 0.135 [0.000, 0.270] & 0.051 & .051\\
& indirect (a$\times$b) & EP+PM $\to$ SA $\to$ Takeover Lead Time & 0.088 [-0.027, 0.203] & 0.037 & .1 \\
\bottomrule
\end{tabular}

\vspace{1mm}
\textnormal{Notes: In Path Type, ``a'' denotes the path from PSAE condition to SA; ``b'' denotes the path from SA to the DV; ``c' '' denotes the direct effect of the PSAE condition on the DV controlling for SA; ``indirect'' denotes the a$\times$b mediated effect. ``Estimation'' reports unstandardized coefficients with bias-corrected bootstrap 95\% CI (N = 5,000 resamples). ``Std. $\beta$'' reports standardized coefficients. Indirect-effect significance was assessed via bootstrap CI (significant if 95\% CI does not include 0).}
\end{table*}

In response to RQ2, the results indicated various relationships between psychological states and neural activity. As shown in Table~\ref{statistic}, higher \textit{Situation Awareness} was a significant predictor of decreased \textit{$P_\alpha$ Power Ratio}, as well as earlier \textit{$P_\alpha$ Activation Time} and a larger \textit{$P_\alpha$ Time Lag}. Higher SA was also significantly associated with an increase in \textit{$F_\theta$ Power Ratio}. Furthermore, higher \textit{Trust} was also linked to a decrease in \textit{$P_\alpha$ Power Ratio}. 

As a foundational step for our subsequent path analysis (RQ3), we found that \textit{Situation Awareness} was a strong predictor of takeover performance. Higher SA was significantly associated with a higher likelihood of a \textit{Takeover Success} and an earlier \textit{Takeover Lead Time}.

\subsection{Path Analysis}

Based on our preliminary regression results, which identified SA as a critical hub, we conducted a formal path analysis to test the mediation mechanism proposed in RQ3. We specified a simple mediation model (\textit{PSAE -> SA -> Performance}). The measurement model for the SA latent variable demonstrated good internal consistency, with all factor loadings exceeding a threshold of  0.5. The path models for both performance metrics achieved a good overall fit \cite{wang2025exploring}.

As shown in Table \ref{path}, the path analysis provided clear evidence that SA acts as a key mediator for the effect of the PSAE interface on takeover success. We found significant indirect effects via SA for all three PSAE conditions when compared to the baseline: for EP (0.431, 95\% CI [0.173, 0.688], $p = <.0001^{*}$), PM (0.327, 95\% CI [0.084, 0.570], $p = .008^{*}$), and EP+PM (0.209, 95\% CI [0.019, 0.400], $p = .03^{*}$). At the same time, the direct effects of the PSAE conditions on Takeover Success (i.e., c' paths) were small and nonsignificant. These results are therefore statistically consistent with full mediation, meaning that PSAE increased takeover success primarily through improvements in drivers’ SA rather than via a direct effect on behavior.

For Takeover Lead Time, the EP condition showed a significant indirect effect via SA (indirect = 0.170, 95\% CI [0.018, 0.321], $p = .03^{*}$). The PM condition demonstrated a borderline indirect effect that approached significance (indirect = 0.135, 95\% CI [0.000, 0.270], $p = .051$), and EP+PM did not show a significant indirect effect. Here, we therefore describe PM as showing a trend-level mediation effect and EP as showing a robust mediation effect.

Furthermore, we conducted a series of path analyses to test the alternative hypothesis that Situation Awareness might also mediate the effect of PSAE on key physiological indicators. However, none of these models achieved an acceptable level of model fit (e.g., all CFIs < 0.89, all RMSEAs > 0.12), indicating a significant discrepancy between our hypothesized causal structure and the observed data. As the models themselves were misspecified, the path coefficients within them cannot be reliably interpreted. This lack of fit suggests that a simple PSAE -> SA -> Physiological causal chain may not be appropriate for our data. The detailed fit indices and parameter estimates for these exploratory but ultimately rejected models are available in the Appendix Table \ref{app:t1}.

\section{Discussion}

\subsection{RQ1: How Interfaces and Environment Shape Driver States}

Our first research question explored the direct impact of the PSAE interfaces and environmental conditions (lighting and hazard visibility) on drivers’ SA, perceived safety, trust, and physiological indicators (EEG/ERSP and EMG measures). The findings confirm that these external factors are influential modulators of drivers' cognitive activities and motor preparation, particularly within the context of a silent failure. A key finding is that the design of the PSAE interface significantly impacts drivers’ trust and perceived safety. The PM and EP+PM interfaces notably increased driver trust, suggesting that providing information about the system’s planned maneuvers is crucial for building confidence \cite{mahadevan2018communicating}.

While abundant information can sometimes cause overload, in our data, the EP interface (displaying environment perception) was associated with significantly higher Situation Awareness compared to Baseline and EP+PM. This suggests that presenting perceptual cues can help drivers better perceive the scene in silent-failure situations. At the same time, the comparatively lower SA in mixed information presentation (EP+PM) suggests that excessive or poorly organized visual information (including AR/highlight displays and dense HUD content) might increase workload and impair SA in driving contexts \cite{gao2022effects}.

Furthermore, we uncovered a "night-time vigilance effect." Drivers in night-time conditions reported significantly higher SA but lower Trust, a finding corroborated by a higher likelihood of muscle preparation. This suggests a compensatory mechanism: when faced with a silent failure under inherently riskier conditions, drivers' reduced trust may compel them to maintain a higher state of vigilance, which aligns with literature on vigilance fluctuations and visibility-related effects on driving performance and preparedness \cite{huang2024enhancing}. This highlights the impact of lighting on drivers’ SA, trust, and motor preparation when handling hazards. Finally, the discovery of a significant interaction effect between PSAE and SA on neural activity (e.g., $F_\theta$ Power Ratio) suggests that the way an interface supports drivers’ cognitive processing during a potential silent failure is not uniform, but is contingent on the driver’s SA level (as measured by SART) \cite{cavanagh2014frontal}.

\subsection{RQ2: Uncovering the Link Between Mind and Body}

Our second research question sought to delineate the relationship between subjective psychological states (SA, perceived safety, trust) and objective physiological indicators derived from EEG and EMG. The most compelling finding here is the consistent and strong association between SA and Parietal Alpha ($P_\alpha$) activity. Our results clearly showed that higher SA was linked to a lower $P_\alpha$ Power Ratio (i.e., alpha suppression) and earlier $P_\alpha$ activation time. This aligns with established cognitive neuroscience literature, where the suppression of parietal alpha power is a well-established neural marker of heightened visuospatial attention and cognitive engagement \cite{woodman2022alpha,deng2019causal}. This convergence validity between subjective and objective measurements has, to a certain extent, strengthened our confidence in using SART scores as a proxy indicator for SA. Meanwhile, our findings empirically ground this theoretical link in a dynamic driving task. This is critical in the context of silent failures, as this neural activity might represent the very first moment the driver's brain flags a problem that the automation might have encountered. This suggests \textit{$P_\alpha$} activity can serve as a potential neural correlate of the implicit subjective SA state. 

Interestingly, while SA was strongly linked to cortical activity (EEG), its direct link to peripheral motor readiness (\textit{Preparation Time}) was not significant. Meanwhile, we found that higher SA was associated with a larger EEG–EMG time lag. This "disconnect" suggests that, while cortical engagement (as indexed by an earlier alpha modulation) marks faster detection/processing of the scene, the increased EEG–EMG lag suggests that this cortical detection does not always translate immediately into peripheral motor commitment. In other words, cognitive detection and motor initiation can be decoupled in silent-failure contexts, which was formally tested in our path analysis. Such a phenomenon echoes previous findings that a good SA does not always mean a good decision or action \cite{nasser2025mismatch}.

We also observed that higher SA tended to go together with increased frontal/theta-band engagement ($F_{\theta}$ Power). Frontal midline theta has been widely interpreted as reflecting cognitive control, conflict monitoring, and the mobilization of executive resources, which are expected when a driver detects a mismatch between the system’s behavior and the environment \cite{cavanagh2014frontal}. Thus, the joint pattern of parietal alpha suppression and frontal theta increase is neurophysiologically coherent: the better SA is associated with enhanced sensory/attentional processing (especially in the case of ADS failures), which facilitates control/monitoring processes \cite{thut2006alpha}.

\subsection{RQ3: The Mediating Role of SART in Driving Performance}

Our third research question investigated the underlying mechanism through which PSAE interfaces affect performance. The path analysis results provide evidence that the effect of the PSAE interface on takeover performance is, to a large extent, indirectly associated with Situation Awareness. Within our experiment, this mediation appears to be an important mechanism for supporting drivers when handling a silent failure: the interface’s primary role may be to facilitate the driver’s self-detection of the problem (by enhancing SA) when the system itself provides no alarm. Prior work on agent transparency and attention-guiding takeover requests similarly suggests that interface transparency and attention guidance improve SA and thereby support takeover performance \cite{van2024agent,chen2023attention} in TOR-initiated takeover scenarios.

Our models for both Takeover Success and Takeover Lead Time converged on this same conclusion. The pattern in the results, with significant indirect effects but without significant direct effects, is consistent with a model in which SA serves as a central pathway through which the interface can influence performance. It is noteworthy that we tested an alternative hypothesis, that SA might mediate the effect of PSAE on physiological states, but found no supporting evidence. This crucial null finding suggests that while physiological states are correlated with driver cognition (as in RQ2), they do not appear to be part of the primary causal chain that flows from the HMI, through SA, to performance in our current models.

This null mediation may reflect several possibilities: a mismatch in the timescales between subjective SA reports and millisecond-level EEG changes, limitations in the signal-to-noise ratio of the physiological measures, or that the true relationship is more complex, perhaps involving conditional effects (i.e., moderated mediation) that our current sample size lacks the power to detect. Future work should use measures with higher temporal resolution to test for potential conditional mediation pathways.

\subsection{Implications}

The collective findings of this study, grounded in the challenging context of silent automation failures, offer several important implications. Our results are consistent with and further support the view that SA is an important construct for safe human-automation interaction \cite{endsley1995measurement}. The mediation pattern observed in response to RQ3 offers one possible explanation in which a key function of transparency-oriented HMIs in silent failures is to enhance the driver’s cognitive understanding of the scenario, which in turn enables effective action. This interpretation is also aligned with the hierarchical model of driving, where tactical-level SA is a prerequisite for operational control \cite{michon1985critical}, and recent theories that treat takeover as a decision process driven by evidence accumulation and risk appraisal \cite{markkula2018models,gonccalves2019applicability,thomas2021uncovering}. 

The consistent relationship between SA and parietal alpha suppression has both theoretical and applied implications. Theoretically, it anchors the abstract construct of SA to an empirically tractable neurophysiological marker (posterior alpha dynamics), strengthening the bridge between neuroergonomics and HMI design. Practically, the consistent relationship between SA and parietal alpha suppression raises the possibility of augmenting HMI systems with neuro-informed monitoring. If further research can confirm that the parietal alpha suppression can reliably index SA in real driving contexts, EEG-derived metrics could potentially be used to support adaptive HMIs that aim at increasing SA. However, three important caveats apply. First, scalp alpha measures can be noisy in real-world settings and require robust preprocessing and individualized baselining \cite{thut2006alpha,deng2019causal}. Second, the temporal relationship we observed between cortical markers and peripheral motor preparation is complex, implying that neurofeedback alone may not be sufficient to predict imminent action. Third, any application that uses neural measures to adapt HMI must be validated for generalizability across drivers, contexts, and device types \cite{wang2025towards}. 

Our findings also provide guidance for the design of in-vehicle interfaces. The results related to RQ1 suggest that more information is not always better. Presenting perceptual information in a clear, salient, and non-competing manner is crucial: poorly organized HMIs may impede drivers' comprehension of the situation. Future HMI designs should not just focus on the type of information to display, but also how to integrate and present the information effectively. For example, we even found a reduced SA under the combined 'EP+PM' interface. This serves as a critical cautionary tale for HMI design: simply adding more information, even if theoretically useful, can lead to cognitive overload or visual interference, ultimately harming performance. Finally, the discovery of the "night-time vigilance effect" highlights that the effectiveness of HMI might be dynamic. This motivates a shift away from "one-size-fits-all" interfaces towards adaptive systems that can tailor information delivery to the specific situation and driver \cite{ding2024one}. The relationship between physiological and cognitive states and behavior also informs the future design of adaptive HMIs. For example, in-vehicle driver monitoring systems \cite{wang2026drowsydg,wang2024multi} can obtain real-time information on the driver’s physiological \cite{wangphysdrive,10903997} and cognitive states \cite{wang2024efficient,wang2026driver} and then flexibly adjust the HMI according to the in- or out-of-cabin context.

\section{Limitations}
We recognize that our study, while following rigorous methods, has several limitations that offer avenues for future research. First, our between-subjects design for the PSAE factor, while necessary to avoid carryover effects, resulted in a small sample size per group. Besides, our regression models show modest but acceptable \(R^2\) \cite{montella2021rule,claveria2019understanding}. We therefore interpret these associations as statistically reliable but modest in size, and do not claim high predictive accuracy at the individual level. These limit the statistical power to detect smaller or more complex interaction effects and necessitate caution when generalizing our findings. Future work could benefit from larger samples or exploring alternative designs. Second, while our high-fidelity driving simulator provided a safe and controllable environment for studying hazardous events, it cannot fully replicate the physical and emotional risks of real-world driving. The psychological and physiological responses observed may differ from those in on-road situations, and future studies should seek to validate our findings in more naturalistic settings.

Further, our path analysis revealed that the proposed physiological mediators have a poor overall fit. While this suggests such simple mediation is not the correct mechanism, it may also reflect several possibilities: a mismatch in the timescales between subjective SA reports and millisecond-level EEG changes, limitations in the signal-to-noise ratio of the physiological measures, or that the true relationship is more complex, perhaps involving conditional effects (i.e., moderated mediation) that our current sample size lacks the power to detect. Moreover, we operationalized takeover performance at the level of immediate response to hazard, i.e., how much lead time the driver had when taking control, without analysing post-takeover driving quality (e.g., lane keeping, speed, or headway control after the intervention). Future work could examine how post-takeover driving performance can be influenced by the visualization of ADS states, which may be affected by different factors through different pathways. Next, our study focused on a specific set of silent failure types and environmental factors. The effectiveness of these PSAE interfaces may vary across different types of hazards, traffic densities, or weather conditions. Other forms of silent failures, such as control-level faults and sudden system shutdowns, were not considered in this study. Future research should explore a wider range of scenarios and silent failure types to build a more comprehensive understanding of how to support drivers in more diverse silent failure scenarios. Finally, we used questionnaire results (e.g., SART, trust and perceived safety) as subjective measures of perceived psychological states. Although we only modeled them as indirect indicators of cognitive processes, it is important to recognize that the self-reported questionnaires reflect participants’ retrospective and overall impressions of each scenario. Such ratings are often influenced by factors like knowledge of the outcome, recall bias, and self-presentation tendencies, and do not offer a moment-to-moment and objective account of drivers’ psychological states during the event. Therefore, the readers should be cautious when interpreting our model and future research may consider objective measures of SA, trust and perceived safety.

\section{Conclusion}
In this study, we conducted a multi-modal driving simulator experiment involving 48 participants to investigate whether different Prospective Situation Awareness Enhancement (PSAE) interfaces can support drivers during safety-critical silent automation failures. By integrating psychological surveys, neurophysiological recordings (EEG, EMG), and behavioral data, we systematically analyzed the pathways from interface design to driver performance. Our analyses, combining mixed-effects regression and path analysis, yielded several key findings:

\begin{itemize}
    \item In our study, Situation Awareness (SA) emerged as a key pathway through which PSAE interfaces were associated with takeover performance. Path analysis revealed a significant indirect effect where the interfaces first influenced SA, which in turn predicted both takeover success and lead time. This highlights that the principal goal for designers of transparency-focused HMIs should be to maximize driver comprehension of the scenario (including both ADS and the traffic environment).
    
    \item Different PSAE interfaces had different effects: EP (perception information) improved SA, while interfaces including planned maneuver information (PM and EP+PM) were only effective in increasing Trust. These results suggest that future in-vehicle HMI designs should consider balancing the richness of the information to increase users' trust in the system while avoiding overloading drivers.
    
    \item We observed patterns consistent with a potential neural correlate of SA. Higher SA scores were associated with suppression of alpha-band activity, earlier neural activation time, and a longer time lag between neural and motor responses. These findings can motivate further work on physiological-informed driver-state monitoring \cite{wang2026drowsydg,wang2024multi} and contextualized and adaptive HMIs, though rigorous field validation with more diverse individuals is still needed.
    
    \item Finally, we observed that the effectiveness of HMI in our study appeared to be context-dependent. The "night-time vigilance effect," where drivers exhibited higher SA but lower trust at night, suggests that external conditions can influence a driver’s internal state. Together with the interaction effects between PSAE and SA, this finding indicates the need for exploring context-aware and adaptive systems that can tailor information delivery to specific situations and heterogeneous drivers.
\end{itemize}

\begin{acks}
This work was supported by the National Natural Science Foundation of China (No. 52202425).
\end{acks}

\bibliographystyle{ACM-Reference-Format}
\bibliography{sample-base}

\appendix

\section{Goodness-of-fit of the SEM models}

\begin{table*}[h] 
 \centering 
 \caption{Goodness-of-fit of the SEM models.} 
 \label{app:t1} 
 \scriptsize 
 \begin{tabularx}{\textwidth}{>{\raggedright\arraybackslash}m{2.6cm} *{9}{>{\centering\arraybackslash}X}} 
 \toprule
 \multirow{2}{*}{\textbf{Goodness-of-fit measures}} & \multicolumn{2}{c}{\textbf{Fit Criteria}} & \multicolumn{7}{c}{\textbf{Model}} \\ 
 \cmidrule(lr){2-3} \cmidrule(lr){4-10}
 & \textbf{Good fit} & \textbf{Acceptable fit} & \textbf{Takeover Success} & \textbf{Takeover Lead Time} & \textbf{$F_\theta$ Activation Time} & \textbf{$P_\alpha$ Activation Time} & \textbf{$F_\theta$ Power Ratio} & \textbf{$P_\alpha$ Power Ratio} & \textbf{$P_\alpha$ Time Lag} \\ 
 \midrule
 Chi-square/Degree of Freedom ($\chi^2$/df) & [0,2] & (2,3) & 1.559 & 3.882 & 3.721 & 3.714 & 3.837 & 3.722 & 3.727 \\ 
 p-value (Chi-square) & < .05 & < .05 & .02 & <.0001 & <.0001 & <.0001 & <.0001 & <.0001 & <.0001 \\ 
 GFI & [0.95,1] & [0.90,0.95) & 0.991 & 0.978 & 0.979 & 0.979 & 0.981 & 0.981 & 0.979 \\ 
 AGFI & [0.90,1] & [0.85,0.90) & 0.983 & 0.954 & 0.957 & 0.957 & 0.961 & 0.961 & 0.958 \\ 
 CFI & [0.97,1] & [0.95,0.97) & 0.979 & 0.973 & 0.889 & 0.888 & 0.884 & 0.888 & 0.888 \\ 
 RMSEA & [0,0.05] & (0.05,0.08] & 0.056 & 0.067 & 0.123 & 0.123 & 0.126 & 0.123 & 0.123 \\ 
 SRMR & [0,0.05] & (0.05,0.08] & 0.061 & 0.052 & 0.05 & 0.05 & 0.051 & 0.05 & 0.051 \\ 
 \bottomrule
 \end{tabularx} 
 \textnormal{\\Notes: GFI: Goodness of Fit Index; AGFI: Adjusted Goodness-of-Fit Index; CFI: Comparative Fit Index; RMSEA: Root Mean Square Error of Approximation; SRMR: Standardized Root Mean Square Residual.} 
 \end{table*}

\end{document}